%% file: main.tex
\documentclass[12pt]{article}
\usepackage[utf8]{inputenc}
\usepackage{geometry} 
\usepackage{authblk}
\usepackage[parfill]{parskip}
\usepackage{multirow}

\input{macros.tex}

\title {Evaluating Automatic Difficulty Estimation \\ of Logic Formalization Exercises}                           

\makeatletter
\newcommand\email[2][]%
   {\newaffiltrue\let\AB@blk@and\AB@pand
      \if\relax#1\relax\def\AB@note{\AB@thenote}\else\def\AB@note{\relax}%
        \setcounter{Maxaffil}{0}\fi
      \begingroup
        \let\protect\@unexpandable@protect
        \def\thanks{\protect\thanks}\def\footnote{\protect\footnote}%
        \@temptokena=\expandafter{\AB@authors}%
        {\def\\{\protect\\\protect\Affilfont}\xdef\AB@temp{#2}}%
         \xdef\AB@authors{\the\@temptokena\AB@las\AB@au@str
         \protect\\[\affilsep]\protect\Affilfont\AB@temp}%
         \gdef\AB@las{}\gdef\AB@au@str{}%
        {\def\\{, \ignorespaces}\xdef\AB@temp{#2}}%
        \@temptokena=\expandafter{\AB@affillist}%
        \xdef\AB@affillist{\the\@temptokena \AB@affilsep
          \AB@affilnote{}\protect\Affilfont\AB@temp}%
      \endgroup
       \let\AB@affilsep\AB@affilsepx
}
\makeatother

\author[1]{Alexandra Mayn}
\author[2]{Kees van Deemter}
\affil[1]{\textit{Department of Language Science and Technology, Saarland University}}
\email{\url{amayn@lst.uni-saarland.de}}
\affil[2]{\textit{Department of Information and Computing Sciences, Utrecht University}}
\email{\url{c.j.vandeemter@uu.nl}}

\date{}
\begin{document}

\maketitle

\begin{abstract}
\noindent
Teaching logic effectively requires an understanding of the factors which cause logic students to struggle. Formalization exercises, which require the student to produce a formula corresponding to the natural language sentence, are a good candidate for scrutiny since they tap into the students’ understanding of various aspects of logic. We correlate the difficulty of formalization exercises predicted by a previously proposed difficulty estimation algorithm with two empirical difficulty measures on the Grade Grinder corpus, which contains student solutions to FOL exercises. We obtain a moderate correlation with both measures, suggesting that the said algorithm indeed taps into important sources of difficulty but leaves a fair amount of variance uncaptured. We conduct an error analysis, closely examining exercises which were misclassified, with the aim of identifying additional sources of difficulty. We identify three additional factors which emerge from the difficulty analysis, namely predicate complexity, pragmatic factors and typicality of the exercises, and discuss the implications of automated difficulty estimation for logic teaching and explainable AI.
\end{abstract}

\section{Introduction}
Mathematical Logic has applications in many domains, from Linguistics to Artificial Intelligence and Engineering. 
% KvD start of addition
Over the last 5 years or so, these applications have been acquiring added importance now that there exists a new focus on making algorithms in Artificial Intelligence transparent. One way in which this has been stated says: ``The principle of transparency requires that any information and communication relating to the processing of those personal data be easily accessible and easy to understand, and that clear and plain language be used" \citep{europe}.

For example, when multi-agent systems construct plans automatically, then the system should be able to ``explain" (i.e., justify) why a certain solution was chosen, and perhaps even allow a user to modify the plan when she disagrees with the system's justification. If the system used logic theorem proving to compute the plan, then the system's explanation will typically involve an explanation of one or more formulas that occur in the proof (\cite{OrenEtAl2020}). For example, the system might choose a particular crane to transport an object because ``all other cranes are either faulty or otherwise occupied", in which case it is this (universally quantified and disjunctive) information that needs to be conveyed to a user. Natural language -- sometimes in combination with graphics -- can offer a good way to convey this information to users, especially if these users are not familiar with Mathematical Logic; in this case Language Technology tools are sometimes used to produce automatic English ``translations" of the formulas involved \citep{alonso}. 

When Mathematical Logic is used -- in philosophy and linguistics, but also in the design of electronic circuits and in certain areas of Artificial Intelligence --
there is often a need for logic teaching, which can be very challenging and time consuming. One type of exercise that is found across logic classrooms is ``formalization" of natural language sentences. In such an exercise, the student gets to see a sentence or text in natural language (e.g., English) and is asked to produce a logic formula that expressed the same information as the sentence.

Mathematical Logic can take many shapes and forms, but its most frequent uses, in Artificial Intelligence and elsewhere,  involve a formalism called First Order Predicate Logic (FOPL). FOPL is also the logic most often taught (e.g., \cite{enderton2001mathematical}, \cite{fitting2012first}). When other logics are involved, these tend to be close variants of FOPL. Formalisation exercises require the student to have a thorough understanding of the meaning of FOPL. Formalisation is an interesting candidate for scrutiny because it taps into the students' understanding of various aspects of logic, and there are a number of studies dedicated to this problem (see \ref{sec:fol_difficulty}).

In order to maximize the effectiveness of logic teaching, it is important to gain an understanding of the factors that can sometimes cause logic students to struggle. Likewise, in Explainable AI, it is important to know where users struggle, because this enables system designers to know what logic formulas are most in need of explanation.\footnote{More subtly, an automated assessment of the difficulty of a formula also allows designers to know what version of a given formula would be the best starting point for producing an English ``translation" (\cite{OrenEtAl2020}).

}
In the present paper we will be engaged with algorithms that attempt to assess the difficulty level of a FOPL formula. The Background Section (Section 2) discusses the work on which our efforts build. Section 3 presents an empirical study that we conducted to evaluate the above-mentioned algorithms. Section 4 discusses the implications of our findings, and the way forward in this challenging area of research.

\section{Background}

A promising approach to assessing both the difficulty of a logic formula and the task of formalising a formula into natural language was proposed in
\cite{perikos2011teaching,perikos2016automatic,perikos2017assistance}, who introduce a tutoring system which is aimed at teaching FOL-formalization to students and providing detailed feedback. Their tutoring system (\cite{perikos2016automatic}) contains a difficulty estimation module based on an elaborate algorithm which takes both logic and natural language features into account. This difficulty module sorts formalization exercises into five difficulty classes, from \textit{very easy} to \textit{advanced}. The authors conduct an evaluation in which they ask three logic tutors to assign a set of test exercises to the five difficulty classes and find that the system's ratings closely match those of human experts, with the F-measure ranging from 0.89 to 1 for each of the five classes. These results appear to be very encouraging; however, we believe that one should be careful when interpreting them. Teachers are not always accurate in assessing exercise difficulty, sometimes overestimating the difficulty of questions which are easy for students, and vice versa (see \cite{van2006teachers} for a review). Therefore, it is worthwhile to investigate how the difficulty predictions of the system relate to an empirical, performance-based metric of difficulty.

We address this question by using data from the well-known Grade Grinder Corpus (\cite{barker2011student}), a resource that contains hundreds of thousands of student submissions to 274 FOL-formalization exercises. In a nutshell, we correlated the difficulty predicted by the system by \cite{perikos2016automatic} with two empirical difficulty measures, proportion of students who solved the exercise correctly on their first attempt, and the average number of attempts before the correct solution. We obtained a moderate correlation with both measures, suggesting that the algorithm used by \cite{perikos2016automatic}'s system indeed taps into important sources of difficulty but leaves a fair amount of variance uncaptured. We then conducted an error analysis, closely examining exercises that were mis-classified, with the aim of identifying additional sources of difficulty. Based on this analysis, we propose a change to \cite{perikos2016automatic}'s way of calculating word order matching and discuss three additional factors which emerge from the difficulty analysis, namely predicate complexity, pragmatic factors and typicality of the exercises.

\subsection{Perikos et al.'s Difficulty Estimation System}
\cite{perikos2016automatic} developed a tutoring system meant to assist university instructors in teaching AI topics, including first order logic. The system can be used for practicing the task of {\em formalizing} sentences into first-order logic; it provides automatic feedback on erroneous solutions. Crucially, it includes a difficulty estimation module, which assigns each formalization exercise to one of five difficulty classes, from \textit{very easy} to \textit{advanced}. The difficulty score is obtained using a set of rules that rely on both natural language features (i.e., features of the English sentence to be formalized) and logic features (i.e., features of a FOL formula that formalizes the sentence). Let's look at the logic features first. 

Logic features include the number of implications, number of quantifiers, the presence of a universal or an existential quantifier, and the number of different connectives. Broadly speaking, the more of every feature there is, the harder the formula is judged to be. \autoref{tab:logic_rule} contains the logic rules. For instance, if the formula in question is $\forall x( Cube(x) \rightarrow \exists y Next(x,y) )$ (``Every cube is next to something"), 

it would be assigned to the class 'Difficult' according to rule 10 because it contains 1 conditional ($\rightarrow$) and both types of quantifiers.

\begin{table}[h]
\begin{tabular}{|l|l|l|l|l|l|l|}
\hline
\multicolumn{1}{|p{1.2cm}|}{\centering rule \\ number} & \multicolumn{1}{|p{1.2cm}|}{\centering num \\ $\rightarrow$} & \multicolumn{1}{|p{1.2cm}|}{\centering num \\ quants} & \multicolumn{1}{|p{1.2cm}|}{\centering $\forall$} & \multicolumn{1}{|p{1.2cm}|}{\centering $\exists$} & 
\multicolumn{1}{|p{2cm}|}{\centering different \\ connectives}&  \multicolumn{1}{|p{1.2cm}|}{\centering difficulty level \\ (output)} \\
\hline
1           & \textless{}2     & 0               & *      & *      & *                     & Very Easy                 \\ \hline
2           & $\geq$2 & 0               & *      & *      & *                     & Easy                      \\ \hline
3           & 0                & 1               & *      & *      & \textless{}3          & Easy                      \\ \hline
4           & 1                & *               & Yes    & No     & $\leq$1         & Easy                      \\ \hline
5           & 1                & \textgreater{}1 & No     & Yes    & $\geq$3      & Medium                    \\ \hline
6           & 0                & \textgreater{}1 & *      & *      & $\geq$3      & Medium                    \\ \hline
7           & 1                & *               & Yes    & No     & 2                     & Medium                    \\ \hline
8           & 1                & $\leq$2   & Yes    & No     & $\geq$3      & Medium                    \\ \hline
9           & $\geq$2 & 1               & *      & *      & \textless{}2          & Medium                    \\ \hline
10          & 1                & *               & Yes    & Yes    & *                     & Difficult                 \\ \hline
11          & 1                & \textgreater{}2 & Yes    & No     & $\geq$3      & Difficult                 \\ \hline
12          & $\geq$2 & 1               & *      & *      & $\geq$2      & Difficult                 \\ \hline
13          & $\geq$2 & \textgreater{}1 & *      & *      & *                     & Advanced                  \\ \hline
\end{tabular}
\caption{The logic-based difficulty rules from \cite{perikos2016automatic}.\\
* stands for ``any number".}
\label{tab:logic_rule}
\end{table}

The logic-based difficulty level is then combined with natural language features to produce the final exercise difficulty level. The natural language features are as listed below. For each of these features, if the feature is found then it adds to the difficulty of the formalization task. 

\begin{itemize}
    \item \textit{Word order matching} penalizes mismatch between the formula and the corresponding sentence and is operationalized as the sum of the distances of the positions of the predicate and of the corresponding word in the natural language sentence. 
    \item \textit{Anaphoric pronouns} (such as "it" or "themselves") are thought to contribute to difficulty because the student needs to map them back to their referents. 
\item \textit{Negation} may lead to confusion because the student needs to decide whether the negated words cancel each other out or not. 
\item \textit{Special words or phrases} are a list of natural language expressions which were collected from the tutors which are argued to cause confusion, e.g. \textit{only} or \textit{unless}. 
\item \textit{Connective mismatch} is based on the idea that certain words normally signal the use of a particular connective, and when that connective is absent from the formula, there is said to be a mismatch. For instance, if the sentence ``Apples and oranges are fruits“ is formalised as $\forall x ((apple(x) \vee orange(x)) \rightarrow fruit(x))$,

then the \textit{and} corresponds to a logical disjunction. 
\item Similarly, \textit{quantifier mismatch} emerges when a certain word invites the use of a particular quantifier but that quantifier is absent from the formula. 
\end{itemize}

\begin{table}[h]
\centering
\begin{tabular}{|l|l|lll|l|l|l|}
\hline
\begin{tabular}[c]{@{}l@{}}FOL\\ diff.\\ level\end{tabular} & \begin{tabular}[c]{@{}l@{}}Word\\ order\\ match-\\ ing\end{tabular} & \multicolumn{1}{l|}{\begin{tabular}[c]{@{}l@{}}Anaph.\\ conns\end{tabular}} & \multicolumn{1}{l|}{\begin{tabular}[c]{@{}l@{}}Neg.\\ words\end{tabular}} & \begin{tabular}[c]{@{}l@{}}Special\\ phrases\end{tabular} & \begin{tabular}[c]{@{}l@{}}Quant.\\ mis-\\ match\end{tabular} & \begin{tabular}[c]{@{}l@{}}Conn.\\ mis-\\ match\end{tabular} & \begin{tabular}[c]{@{}l@{}}Diff. level\\ (Output)\end{tabular} \\ \hline
Easy                                                        & 0                                                                   & \multicolumn{1}{l|}{\textgreater{}1}                                        & \multicolumn{1}{l|}{*}                                                    & *                                                         & yes                                                           & *                                                            & Medium                                                         \\ \hline
Medium                                                      & *                                                                   & \multicolumn{1}{l|}{*}                                                      & \multicolumn{1}{l|}{\textgreater{}1}                                      & *                                                         & *                                                             & *                                                            & Difficult                                                      \\ \hline
Difficult                                                   & *                                                                   & \multicolumn{3}{l|}{SUM \textgreater{}3}                                                                                                                                                                            & *                                                             & *                                                            & Advanced                                                       \\ \hline
\end{tabular}
\caption{A subset of natural language feature-based difficulty rules from \cite{perikos2016automatic}. * stands for ``any number". See the original paper for the whole table.}
\end{table}

An important aspect of the proposed algorithm is thus that, given an English sentence in need of formalisation, it requires a reference formula to compute the logic features, as well as some of the language features. For example, when we discussed the above sentence ``Apples and oranges are fruits", we assumed that it has the reference formula 
$\forall x ((apple(x) \vee orange(x)) \rightarrow fruit(x))$ and concluded there was a connective mismatch. But if the sentence has the reference formula $\forall x (apple(x) \rightarrow fruit(x)) \wedge \forall x (orange(x) \rightarrow fruit(x))$, there is no such mismatch (though a word order mismatch appears in its place). In theory, there is an infinite number of equivalent formulas which can be used as the reference, potentially leading to a different difficulty score. The authors do not discuss the question of how to choose a reference formula for a given English sentence; our own investigations have used a corpus-based approach to this issue (see Section 3.1).

The authors conducted an evaluation of their system by asking three tutors to assign a difficulty level to a set of test sentences and correlating the tutor ratings with the system predictions. The system’s predictions closely matched the difficulty levels assigned by students, with the F-measure of the 5 classes ranging between 0.89 and 1. However, the fact that the system behaves similarly to the tutors does not necessarily mean that it makes accurate difficulty predictions. Speaking in general, there is evidence that teachers often give an inaccurate difficulty assessment to an exercise \citep{van2006teachers}. 
Moreover, since the rules of this particular system were obtained from logic teachers,\footnote{The authors write that the rules were obtained by consulting ``expert tutors in the field from our university having many years' experience of teaching AI and logic courses".} it is perhaps unsurprising that logic teachers find them plausible. We therefore believe that another, additional type of evaluation is called for.

\subsection{Grade Grinder Corpus}
The Grade Grinder Corpus (\cite{barker2011student}) is a collection of student submissions to 274 FOL-formalization exercises. It includes over 4.5 million submissions by 55,000 students, which were submitted to the automatic grading system, Grade Grinder, as part of a logic course between 2001 and 2010. Of those exercises, 70\% of students’ first attempts, and 83.4\% of all attempts, are correct. Most of the remaining attempts are incorrect, but other answer types exist, such as missing and ill-formed answers. The exercises come from three domains. 222 of the 274 sentences are from the so-called \textit{Tarski's World} domain, which describes simple stylised objects, their geometric and size properties and their relationship to each other. Of the remaining sentences, 42 come from the \textit{pets} domain, which describes pets and their owners, and 10 come from the \textit{number} domain, which describes properties of numbers. Additionally, while the main task for all the exercises is formalization of sentences in FOL, the exact task type varies in the corpus. For instance, for some exercises, the students look at the picture of the world while formalizing them. For others, a skeleton translation is given and the student’s job is to fill in the missing parts. This is important to keep in mind since the exact task type might be a factor in exercise difficulty. 

\begin{table}[]
\centering
\begin{tabular}{llrr}
\hline
                      & avg (sd)    & \multicolumn{1}{l}{min} & \multicolumn{1}{l}{max} \\
                      \hline
quantifiers           & 1.29 (1.27) & 0                       & 7                       \\
connectives           & 2.27 (2.67) & 0                       & 21                      \\
different connectives & 1.78 (0.93) & 0                       & 4                       \\
predicate arguments   & 1.62 (0.57) & 1                       & 4        \\
\hline
\end{tabular}
\caption{Statistics relevant to the complexity of formulas contained in the Grade Grinder corpus.}
\label{tab:summary_stats}
\end{table}

Some summary statistics regarding the complexity of the formulas contained in the corpus can be found in \autoref{tab:summary_stats}.

\subsection{Difficulty of logic and maths problems}

Many studies have sought to understand the relation between human reasoning, which is resource bounded and involves all kinds of human misunderstandings and biases, and formal reasoning, which tends to rest on the ``normative" idea of validity in mathematical logic (e.g. \citep{johnson1983mental}). Here we briefly review strands of this work that are relevant to our present endeavour, starting with studies that focussed on formalisation from Natural Language sentences.

\subsubsection{Difficulty of logic formalization} \label{sec:fol_difficulty}
Our own work is most closely related to studies that used the Grade-Grinder corpus to investigate factors that might contribute to formalization exercise difficulty. 

\cite{barker2008empirical} conducted a detailed analysis of incorrect formalisations that students had proposed for
20 sentences. The most frequent errors turned out to be antecedent-consequent reversal (where the natural language sentence confuses the antecedent with the consequent), and substitution of the conditional for the bi-conditional. A follow-up study (\cite{barker2009dimensions}) considered the same set of 20 sentences and examined them in terms of two empirical measures of difficulty, namely the proportion of students who submitted an incorrect solution on their first attempt and the number of attempts it took students to correct an erroneous solution. They showed that, while correlated, these two measures do not tap into exactly the same concept because there are sentences which are easy to get wrong but easy to fix, and the other way around. \cite{barker2011impedance} studied the influence of type of information on difficulty and found that the interaction of visual and spatial features affects exercise difficulty. For instance, they showed that sentences containing spatial information were more difficult to formalize than sentences containing information about size and shape. 

All of these studies were exploratory in nature. They did not involve algorithms for automatically quantifying the difficulty of a formalisation exercise. They also only look at a small portion of the Grade Grinder corpus. In the present study, we adopt the metrics used by \cite{barker2009dimensions} but we use the entire Grade Grinder corpus.

\subsubsection{Difficulty of other types of problems}
While difficulty of formalization exercises has been addressed by a number of studies, it is worthwhile to also examine the literature which looks at complexity estimation of other logic and math problems. 

Some studies in this category still focussed on mathematical logic (though not on formalisation). In many cases, the focus was Description Logic, a class of formalisms often used for encoding formal ontologies (e.g. \cite{baader2017introduction})
It was found that subjects have trouble distinguishing formulas of the form $\neg \exists x(F(x))$ (``No object has $F$") from those of the form $\exists x (\neg F(x))$ (``Some objects do not have $F$") \cite{rector2004owl}, where the scope of the negation changes the meaning. The same authors also identified problems with Description Logic's use of the $\forall$ quantifier. For example, the Description Logic construct $\forall ParentOf.Female$
denotes the set of things 
such that {\em all} things they are parents of are female (i.e., the set of people who have no sons). This construct is often misunderstood as requiring the elements of this set to have at least one daughter (i.e., with existential import). These issues, and ways to address them by modifying the Description Logic formalism, have been studied further in \cite{warren2019improving}, where a useful overview of this area is offered.

\cite{strannegaard2013reasoning} conducted a study in which participants were asked to verify whether a given a first-order logic formula was true or false given a certain model. They correlated participants' response latency and accuracy with a number of logical as well as cognitive difficulty measures. Length of the proof needed to determine truth or falsity had the highest correlation with those measures, followed by working memory capacity.

\cite{newstead2006predicting} modeled the difficulty of problems from the reasoning section of the Graduate Record Examination (GRE), a test required for admission to many graduate schools in the United States and Canada.
Their approach relied on variables like the complexity of the used rules, as well as the number of mental models (\cite{johnson1983mental}) required to represent the problem. They first conducted a pilot study to identify variables that should be included in their model, and then validated their model in a larger-scale experiment.

In order to identify factors contributing to the difficulty of mathematical problems, the studies discussed above define models and then correlate their predictions with empirical measures of difficulty, e.g. accuracy and latency the case of \cite{strannegaard2013reasoning}. In our study, we apply a very similar method to empirically validate the sources of difficulty in logic formalization exercises that \cite{perikos2016automatic}'s algorithm relies on.

\section{The present study}
In this study, we correlate the difficulty scores assigned to the Grade Grinder exercises by \cite{perikos2016automatic}’s system with two measures of student performance. We then conduct an error analysis with the goal of identifying other important factors which contribute to exercise difficulty. 

\subsection{Evaluation using the Grade Grinder Corpus}
Perikos’s rules do not cover the connective  $\leftrightarrow$ (``if and only if", also known as equivalence), and the authors told us that their tutoring system did not contain any formalization problems with that connective, so we first filter out the 7 sentences which contain equivalence from the corpus. 

For the remaining sentences, we calculate two empirical difficulty measures - the proportion of students who solved the exercise correctly on their first attempt (First Attempt Correct, FAC) and the average number of attempts among all students preceding the correct answer (Avg. Attempts, AA). We saw that AA included outliers with more than 26 attempts for some sentences, so in our calculation of AA we discarded submissions which were beyond three standard deviations from the mean number of attempts for each sentence. FAC is a real number between 0 and 1, AA is a real number between 1 and 6.11. These measures are quite similar to those used in \cite{barker2009dimensions}. Our two measures are very strongly correlated with each other (\textit{r=}-0.85, \textit{p\textless}0.0001), supporting the intuition that a more difficult sentence will as a rule be solved by fewer students correctly the first time and will also take more attempts to correct. However, we decided to keep both of these measures since they appear to not always tap into the same concept (see \cite{barker2009dimensions} for discussion). 

We assigned a difficulty score to each exercise following the procedure outlined in \cite{perikos2016automatic}. For assigning the difficulty score, a reference formula is required (see section 2.1). Since there are many equivalent correct answers, some of which might have different difficulty scores, we computed the difficulty score for each submission and averaged those individual scores to obtain a final difficulty score for the sentence. The difficulty score is thus a real number between 0 and 4.\footnote{Alternatively, we could have taken the difficulty of the most preferred solution or the minimum difficulty. 

Experimentation suggested that a change to these alternative metrics would only minimally affect the scores. Therefore, we decided not to complicate matters unnecessarily, proceeding as outlined above.} 

We then correlated the difficulty scores with both performance metrics. The results are presented in Table 2. For all sentences, the difficulty level assigned by \cite{perikos2016automatic} is significantly moderately correlated with the performance metrics, with a stronger correlation with FAC, \textit{r=}-0.42 than with AA, \textit{r=}0.33. We hypothesized that task type might be a confounding factor, so we also calculated the same correlations for sentences in the two most common task types, \textit{looking at world} and \textit{with world check}. Indeed, the correlations with Perikos’s scores are stronger in this case, up to \textit{r=}-0.58 for FAC and \textit{r=}0.51 for AA for the \textit{looking at world} task type. 

We also calculated the correlation of the difficulty level assigned by Perikos's system just based on the logic features, and for most conditions, the correlation was significantly worse (e.g. for all sentences, it was -0.37 and 0.23 for FAC and AA respectively). One interesting exception is the \textit{looking at the world} subcondition, where the students looked at the picture describing the world in which the sentence was true while translating it: for that subcondition, using language-based features on top of the logic-based ones improved the correlation with AA (0.27 vs. 0.51, \textit{t=}10.72, \textit{p=}0 on a two-tailed paired correlations test) but not with FAC (-0.55 vs. -0.59, $p>$ 0.1). It is not quite clear how to interpret this result; it might suggest that at least in some cases, the natural language phrasing of the exercise matters for how likely someone is to fix it once they got it wrong, but not so much for getting it correct the first time. In general, however, we observe a significant contribution of the NL features.

\begin{table}[h!]
\centering
\begin{tabular}{llll}
\hline
exercises        & num sents & FAC   & AA   \\
\hline
all              & 274       & -0.42 & 0.33 \\
looking at world & 90        & -0.58 & 0.46 \\
world check      & 87        & -0.53 & 0.51 \\
\hline
\end{tabular}
\caption{Correlation of the average difficulty score assigned by \cite{perikos2016automatic}'s algorithm with empirical measures of difficulty, \textit{first attempt correct} (FAC) and \text{average number of attempts} (AA) for the sentences in the Grade Grinder Corpus. All correlations are significant at \textit{p\textless}0.0001.}
\end{table}

\begin{figure}[htp]
\includegraphics[width=.5\textwidth]{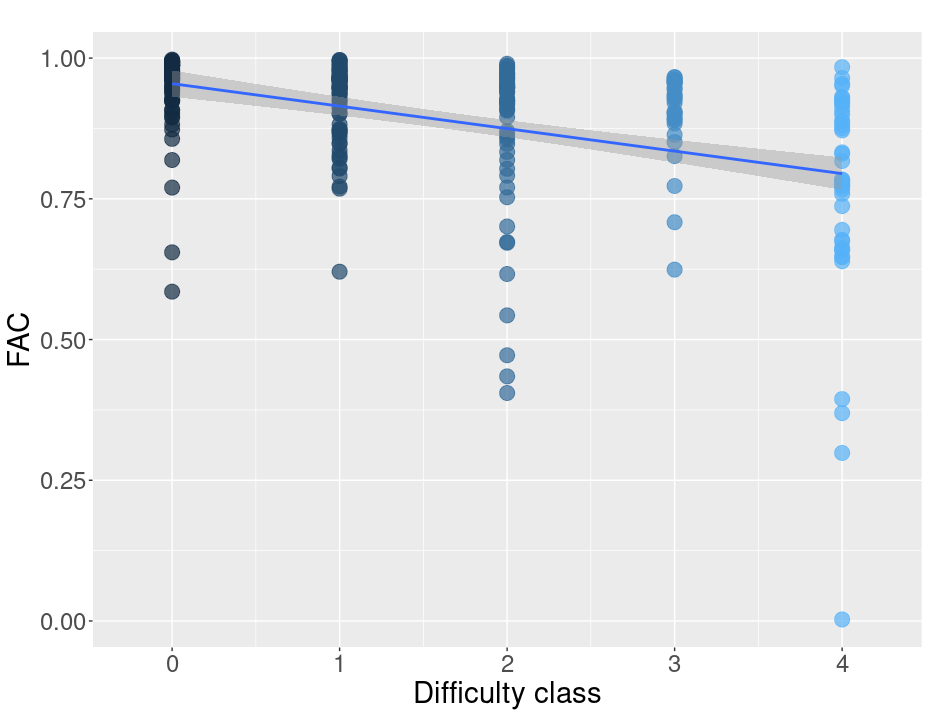}\hfill
\includegraphics[width=.5\textwidth]{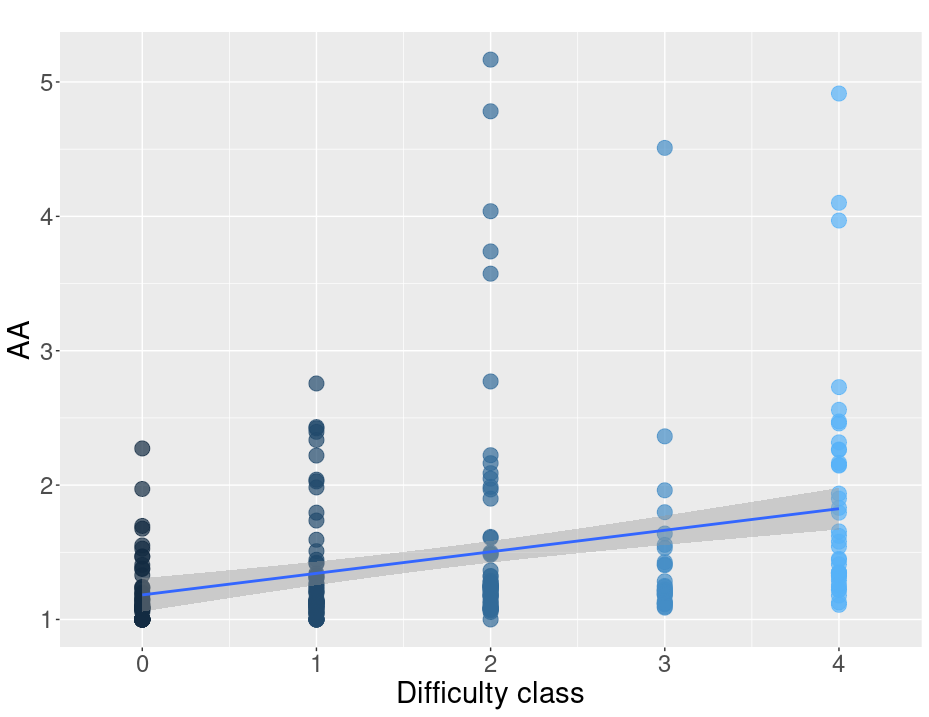}\hfill
\caption{First attempt correct (left) and average number of attempts (right) vs. the difficulty score assigned by \cite{perikos2016automatic}'s system for all sentences. A general downward trend for FAC and an upward trend for AA (more difficult sentences are solved correctly the first time less often and take more attempts to solve) is apparent but the spread, especially around the difficulty classes 2 and 4, is quite large.}
\label{fig:FAC_vs_diff_class}
\end{figure}

These results suggest that, at least on the Gradegrinder data\footnote{See Section 4.2 for a discussion of this qualification.}, the difficulty estimation criteria proposed by \cite{perikos2016automatic}. capture some important sources of difficulty. However, it also serves to make the point that expert evaluation does not always provide an accurate prediction of student performance. 

We were interested to know which of the features in \cite{perikos2016automatic}’s algorithm have the biggest weight in determining difficulty. We trained a machine learning Random Forest-based regressor and used it to predict FAC and AA based on all 11 features in \cite{perikos2016automatic}. We then extracted feature importances, and observed that {\em number of quantifiers} was the feature that the regressor\footnote{Results were very similar with XGBoost and Extra Trees.} assigned by far the most weight to, 0.34 and 0.58 out of 1 for FAC and AA respectively. Word order matching was second in importance, with the weights of 0.23 and 0.14 respectively for the two measures. The rest of the features were assigned quite small weights. We then correlated the (average) number of quantifiers (grouped into 5 levels using 1-dimensional k-means clustering in order to be comparable with \cite{perikos2016automatic}’s  difficulty scores) for each submission with FAC and AA and obtained the scores of \textit{r=}-0.4 and \textit{r=}0.32 respectively, which is comparable to and not significantly different from the correlations of those empirical metrics with \cite{perikos2016automatic}’s difficulty score. It is important to note that the number of quantifiers and \cite{perikos2016automatic}’s final difficulty class assignments also correlate very highly with each other at \textit{r=}0.87.

\subsection{Error analysis}
Having seen that \cite{perikos2016automatic}’s difficulty scores have a moderate correlation with the student performance metrics, we now examine concrete sentences which were misclassified more closely. We are interested in two cases: those which were assigned a high difficulty score but were actually easy to solve, as evidenced by a high FA and a low AA; and those which were classified as easy but in practice caused students difficulty, resulting in a low FA and a high AA.

\subsubsection{Predicted difficult, actually easy}
One example where the system’s prediction differed from the observed difficulty quite strikingly was Exercise 14.4.2, “There are exactly 3 cubes”. The average difficulty level assigned by the system was 3.73 out of 4. However, the students appeared to have very little trouble with it, with 93\% of first attempts being correct and an average of only 1.13 attempts. 

A reason might be familiarity: logic courses often pay much attention to formulas of this kind, because of the obvious importance that numbers (such as 3, for instance) have in mathematics. In fact the LPL textbook (\cite{barwise2000language}), which is used for the logic courses which employ Grade Grinder, contains an example solution of an extremely similar formalization exercise, “There are exactly 2 cubes”. To firm up this idea, we note that it has been shown in other areas of cognition that familiarity, even at the symbol level, has a positive effect on performance, and this positive effect increases with task difficulty (\cite{shen2020icon}).

Another telling example for which the difficulty was overestimated by the system is Exercise 11.16.4, “A large cube is in front of a small cube”. The most frequent correct solution is  $\exists x (Cube(x) \land Large(x) \land \exists y (Cube(y) \land Small(y) \land FrontOf(x,y)))$, which is assigned the highest difficulty class by \cite{perikos2016automatic} because the order of predicates in the formula does not match the word order of the sentence. However, people don’t appear to have any difficulty with the order of conjuncts - indeed, swapping them in any way would result in an equivalent formula which would possibly be assigned a lower difficulty score by the system. We therefore hypothesized that the order of conjuncts and disjuncts shouldn't matter for the total mismatch, but mismatches for non-transitive connectives, e.g. antecedent-consequent reversal, should. We therefore implemented an alternative version of the \textit{word order matching} function which did not overcount such conjunct and disjunct mismatches. However, that did not result in a significant improvement in correlation with either FAC or AA, presumably because the contribution of \textit{word order matching} to \cite{perikos2016automatic}'s final difficulty score calculation is so small, for example compared to $num\_quantifiers$.

\subsubsection{Predicted easy, actually difficult}
We observed that there were fewer such cases - difficulty seems to be over- rather than underestimated by the system.

An interesting example is Sentence 7.18.2, from the \textit{Pet} domain. The sentence to be formalized is as follows: “Max fed Folly at 2pm, but if he gave her to Claire then, Folly was not hungry five minutes later.” This translates to the following formula (in addition to other, logically equivalent answers): 

$Fed(M,F,2:00) \land (Gave(M,F,C,2:00) \rightarrow \neg Hungry(F,2:05))$.

\cite{perikos2016automatic}’s system assigns this sentence to the easiest difficulty class. However, only 65\% of the students solve it correctly the first time and 2.27 attempts are needed on average. What are the sources of difficulty which aren’t captured? To answer that question, we look at the most frequent incorrect solutions, and the one that stands out and accounts for 19\% of the 1214 incorrect solutions is the following: 

$(Fed(M,F,2:00) \land Gave(M,F,C,2:00)) \rightarrow \neg Hungry(F,2:05)$.

There is a bracketing error here, where the predicates representing feeding and giving are grouped together.

We believe that this is an error caused by pragmatics: world knowledge tells us that being hungry is related to feeding, not giving. So this error was supposedly made in an attempt to make sense of a pragmatically improbable sentence. This illustrates pragmatic effects on FOL-formalization. Such effects would be quite difficult to capture automatically but could be considered by the instructor when creating or assigning exercises.

Another factor which might be at play here is readability. The sentence about Folly is probably quite difficult to read. \cite{barker2011impedance} found an effect of readability and informational complexity, which \cite{perikos2016automatic} appears to account for only partially through features like anaphoric pronouns and negation words. 

We also see that the predicates in this exercise contain many arguments, 
indicating a potentially more complex relation between arguments. We hypothesized that the number of arguments in the predicates might be an important predictive feature. To test this hypothesis, we fitted two sets of linear models - one with just the difficulty class assigned by \cite{perikos2016automatic}'s system as the predictor variable, and another where the average number of predicate arguments in the solution is also included as a predictor. In each set, the dependent variable is one of our empirical difficulty measures - FAC and AA. One-way ANOVA indicated that in both cases, adding the number of predicate arguments as a predictor significantly improved model fit (F(1) = 33.7, \textit{p \textless} 0.0001 for FAC and F(1)=12.1, \textit{p \textless} 0.001 for AA). This suggests that the complexity of predicates in the formula is a meaningful predictor of difficulty, which would lead to more accurate difficulty estimations if incorporated in an automated estimation system like that of \cite{perikos2016automatic}.

\section{Discussion}
\subsection{Discussion of the findings}
In this study, we evaluated the complexity estimation algorithm for logic formalization exercises proposed by \cite{perikos2016automatic} using two metrics of student performance on the Grade Grinder exercises (\cite{barker2011student}), proportion of students whose first attempt at the exercise was correct (FAC), and the average number of attempts needed to solve the exercise correctly (AA). We then conducted an error analysis in order to identify additional sources of complexity that the algorithm proposed by \cite{perikos2016automatic} didn't account for.

The assigned difficulty classes correlated moderately with the empirical difficulty metrics, with the highest correlations in the 0.5-0.6 range for subtasks of the same type. These results can be regarded as quite good, yet they are very different from those obtained using the expert evaluation reported in the paper (F1-score in the range of 0.89 and 1). This shows that caution is in order when using only expert evaluations because experts are not always able to accurately assess how difficult the material will be for learners, and that an expert evaluation should be accompanied by one with the actual target audience of the material.

Difficulty is a complex notion that can be defined in different ways, tapping into potentially different underlying concepts: 
we see that our metrics, FAC and AA, correlate quite strongly but not perfectly with each other (\textit{r=}0.85) and moderately (\textit{r=}-0.41 to -0.59 for FAC and \textit{r=}0.32 to 0.51 for AA) with the difficulty scores assigned by \cite{perikos2016automatic}'s algorithm. Further investigation into what those underlying concepts are could be essential to creating better difficulty estimation algorithms.

\cite{perikos2016automatic}'s difficulty estimation system does identify some features that, taken collectively, contribute to difficulty, as indicated by their generally quite good correlation with the empirical performance metrics. Interestingly, however, most of the features in their algorithm appear not to contribute much: one feature, namely the number of quantifiers, appears to be by far the most important one, as evidenced by the fact that the correlation of the empirical metrics with just that one feature grouped into five levels is not significantly different from the correlation of those metrics with \cite{perikos2016automatic}'s final score using all features. 

Our error analysis revealed some other interesting factors contributing to complexity. First, we saw that, like in language comprehension more generally (\cite{grice1975logic}), pragmatic factors play a role: if the sentence to be formalized is pragmatically improbable, the learner might make an error trying to make sense of that sentence. This idea would be very challenging to operationalize and to measure automatically; however, it could be borne in mind when developing or selecting exercises. Second, we saw that familiarity is a factor that influences difficulty: notably, students had no trouble with an exercise that was assigned the highest difficulty class by \cite{perikos2016automatic}; closer investigation revealed that a practically identical exercise was explained in the chapter of the book used by the students in the course. A familiarity metric could be incorporated into the algorithm for determining difficulty by, for instance, comparing the exercise to those in the textbook and computing a similarity score. Finally, we saw that the average number of arguments of the predicates in the formula was a significant predictor of difficulty which improved on the system's predictions, presumably because a larger number of predicates signals a more complex relation, which would also translate into a more complicated sentence.

Existing work has started to explore the relationship between the computational complexity of quantified logical formulas and the cognitive difficulty humans experience when processing those formulas. For example, \cite{mcmillan2005neural} showed that different brain regions are involved in processing sentences containing quantifiers which are recognized by finite state automata (FA), versus push-down automata (PDA), higher-order quantifiers. In a similar vein, \cite{szymanik2010comprehension} examined a more fine-grained quantifier subgrouping, that of quantifiers recognized by acyclic FAs, FAs with loops and PDAs, and found that reaction times of humans comprehending quantified sentences were significantly higher for quantifiers recognized by more complex automata. For example, sentences with the quantifier "most" is harder than ones with the quantifier "three", and these were harder than ones with the quantifier "at least one").

We asked ourselves whether a similar pattern would emerge here - whether students would have less difficulty with exercises containing monadic formulas (i.e., formulas all of whose predicates have only one argument place) than with one that involve formulas of higher arity. Unfortunately, in the Grade Grinder corpus, there is only a very small number of monadic quantifiers (35 out of the 265 sentences we considered), and all of those are very simple along other dimensions (they all have only one quantifier and most of them have only 2 predicates), which makes them not directly comparable to the full FOL sentences. It would be interesting to examine whether exercises that correspond to monadic formulas tend to have different difficulty levels than other formulas when formulas are matched for length, for the number of predicates, and for the number of quantifiers. We imagine that that is likely to be the case, since the average number of predicate arguments in a formula turned out to be a significant feature in predicting exercise difficulty in our error analysis.

There are other possible sources of difficulty one could speculate about, which \cite{perikos2016automatic}'s algorithm doesn't consider - one is the total number of different predicates: we suspect that a sentence like "There are only three cubes" should be no more difficult than "There is only one cube", but because the former's formalisation requires more quantifiers (since each one of the cubes has to be enumerated separately in the formula), it is nonetheless assigned a higher complexity. 

Another factor that could be considered is symmetry. The basic intuition here is that when a formula repeats the same syntactic construct $n$ times in a row, then the formula is easier than a similar formula that mixes different very syntactic constructs. For example, a formula of the form $\forall x F_1(x)$ $\&$ $\forall x F_2(x)$ $\&$ ... $\&$ $\forall x F_n(x)$ should not be penalized for containing $n$ universal quantifiers, because this happens in a uniform sequence of conjuncts, each of which is almost exactly like the others, and hence not adding much difficulty. Another possibility, which would have similar effects in the case of this formula, 
would be to assess the difficulty levels of unrelated conjuncts separately, and to legislate that the difficulty of such a conjunction is equal to the highest difficulty level that is found in any of its conjuncts, instead of adding up the difficulty levels of the conjuncts; the same could be done for disjuncts. This would be a drastic change to Perikos' approach, with huge effects particularly on longer formulas (which happen to be rare in the Gradegrinder corpus). 

Needless to say, these ideas would have to be tested empirically.

\subsection{Limitations and future work}
Finally, let's reflect on the enterprise itself of designing and evaluating difficulty estimation algorithms for logic formulas. Regardless of what predicates and arguments are permitted, and of how many arguments a predicate can have, the class of FOL formulas is mathematically infinite.\footnote{This can easily be seen from the recursive structure of the syntax rules for FOL. For example, the rule for negation says that if $\varphi$ is any formula, then $\neg\varphi$ is likewise a formula. Given one atomic formula, e.g., Even(4) (``The number 4 is even"), this rule alone yields all the formulas  $\neg Even(4)$, $\neg\neg Even(4)$, $\neg\neg\neg Even(4)$, ... The rules governing connectives and quantifiers operate in similar ways.} What does it mean to propose an algorithm that estimates the difficulty level of each member of this infinite class? And even more crucially, how should formulas be sampled from this infinite space?

It is important to acknowledge that difficulty estimation is always relative to a task. The task for which the formulas are used determines the class of formulas on which the algorithm and its evaluation should focus. FOL-Formalization is an important task students will need to master to gain an understanding of First-Order Logic, but it is by no means the only one. For example, in the type of ``Explainable AI" discussed in the Introduction, the skill of being able to interpret a formula (and possibly render it in English -- going in the direction opposite to formalisation) might be at least as important to master. Other important tasks include verification (i.e., Is a given formula true with respect to a given domain?), inference (Does one formula follow from another?), and so on.

It is worth pondering how much and what kind of adaptation would be required to extend an algorithm such as the one proposed by \cite{perikos2016automatic} to a different task or to a larger subset of FOL-formulas. By definition, formulas that students produce when formalizing natural language sentences are human-authored and, therefore, not include formulas of unintuitive structure, such as formulas that are extremely lengthy, or that use very deep levels of embedding. Yet such formulas abound in other areas. 

Consider, for example, formulas that contain vacuous quantifiers (i.e., quantifiers that do not bind any variables). A simple example is $\forall x Even(4)$, which is a well-formed formula of FOL even though $\forall$ does not bind any variables. An algorithm like that of \cite{perikos2016automatic} would assign this formula a low difficulty level, since it contains just one quantifier and one predicate, yet students struggle, because they find it difficult to grasp the effect of a quantifier that fails to bind any variables.\footnote{In fact, such quantifiers do not change the meaning of the formula: the formula above means the same as $Even(4)$.} 

It is worth seeing why formulas that are not ``well-behaved" can nonetheless be of considerable importance in logic teaching. 
For example, such formulas are routinely produced by any automated theorem prover, for example when the prover converts a premiss into Skolem Normal Form (e.g., \cite{robinson-voronkov}); such normal forms were not constructed for human consumption but merely as intermediate steps in a formal proof. Making such formulas understandable to people can be important in the context of applied multi-agent systems, for example, where theorem provers become much more ``scrutable" (i.e., transparent) when the formulas that they use can be grasped by human users \cite{OrenEtAl2020}.

Another class of formulas that are well formed yet not ``well behaved" are the formulas employed by semantic parsers in the tradition of Richard Montague (e.g., \cite{popescu-etal-2003}). In a nutshell, semantic parsers compose a formula $\phi$ that represents the meaning of a sentence $S$ from a number of formulas, say $phi_1,..,\phi_n$, that represent the meanings of the $n$ syntactic parts of $S$. Even if the formula $\phi$ is relatively simple, $phi_1,..,\phi_n$ can often be highly complex (e.g. containing numerous $\lambda$ operators), making it difficult for builders of semantic parsers to grasp their meaning, and this complexity motivates a considerable amount of logic teaching.\footnote{Although the semantic representations employed by semantic parsers go beyond FOL (e.g., by including $\lambda$ and modal operators), they tend to have FOL at their core.}

These examples illustrate that when algorithms such as that of Perikos and colleagues, which claim to assess the difficulty of logical formulas are proposed and evaluated, it is not enough to specify the formalism (e.g. FOL). Crucially, the specific class of formulas the algorithm applies to -- which is always only a small subclass of the set of all formulas permitted by the formalism -- should be made explicit. This could be done, for example, by specifying a maximum formula length and adding further constraints such as a prohibition against vacuous quantifiers.

The discussion above also demonstrates the diversity of the beneficiaries of the algorithms discussed, who vary from engineers to linguists and people who work on automated theorem proving. For all these reasons, we believe that it is important that future work on algorithms for estimating the difficulty of logical formulas should be as specific as possible about the aims and purposes of the work and, consequently, not only about what type of learners they want to target, but also about the precise class of formulas whose difficulty the algorithm is intended to assess. 

\bibliography{main}
\bibliographystyle{acl}

\end{document}

%% file: macros.tex
\usepackage{natbib}
\usepackage{xcolor}
\usepackage{url}
\usepackage{hyperref}
\usepackage{mathtools}
\usepackage{booktabs}

\usepackage{algpseudocode}%[noend]
\usepackage{algorithm}

\usepackage{float}
\usepackage{yfonts}

\usepackage{amscd}
\usepackage{amsfonts}
\usepackage{amsmath}
\usepackage{amssymb}
\usepackage{enumerate}
\usepackage{graphicx}
\usepackage{mathrsfs}	
\usepackage{wrapfig}
\usepackage{amsthm}

\usepackage{mathrsfs}

\usepackage{subcaption}
\usepackage{xspace}

\usepackage{ stmaryrd }